\documentstyle[psfig,floats,aps,eqsecnum] {revtex}
\begin{document}
\newcommand{\vel}{{\bf v}}
% The following line is crucial to get two-column format right
\twocolumn[\hsize\textwidth\columnwidth\hsize\csname @twocolumnfalse\endcsname
\title{ Non-fermi liquid as passive scalar fluid }
\author{Jonathan Miller}
\address{NEC Research Institute, Princeton, NJ 08540}
%\date{{\today}}
\maketitle

\begin{abstract}

I suggest that electron localization by random flux and passive transport
in quenched velocity fields in two dimensions
be studied as perturbations of the simple operator ${\cal K}={\bf A} \cdot \nabla$,
with incompressible velocity field/vector potential
${\bf A}=\nabla \times \phi=(-\partial_y,\partial_x)\phi$.
This operator has an infinitely degenerate subspace of
zero energy eigenstates, arising from incompressibility, that
are {\it extended} for generic $\phi({\bf x})$ and are expected to remain so
under perturbation. I propose that an anomaly accounts qualitatively
for properties of the spectrum and eigenstates of ${\cal K}$ and
its perturbations.
\end{abstract}
\vskip2pc]
\narrowtext

In two dimensions, all eigenstates of the hermitian
operator ${\cal H}_V = -\nabla^2 + V({\bf x})$
are localized for random $V$, no matter how small the randomness
(Anderson localization).
Over the years it has emerged that certain kinds of fields
can suppress localization. In particular,
a strong {\it uniform} magnetic field
$B \! = \! \nabla \! \times \! {\bf A} \! = \!
\partial_x A_y \! - \! \partial_y A_x$ produces
a dramatic change in the character of the spectrum of
${\cal H}_B = (-i\nabla \! - \! {\bf A})^2/2m$, 
yielding Landau levels consisting of degenerate extended eigenstates
that remain extended on perturbation by random $V$.

The underlying physics of this phenomenon,
the quantum Hall effect, can be thought of
as arising from incompressibility. In brief, 
taking $m \to 0$ in the Lagrangian
${\cal L}_B = m{\dot{x}^2}/2 + {\bf A} \! \cdot \! \dot{\bf x}$,
one obtains the action
$\oint{\! dt \, {\bf A} \! \cdot \! \dot{\bf x}  } =
 \int_{\Omega}{ \! d^2 \! a \, B }$.
This action is manifestly invariant
under area-preserving deformations of $\Omega$. The invariance
persists after quantization - albeit in a different form - and is the
origin of \lq quantum incompressibility\rq \cite{trugenberger}.

This version of incompressibilty is expected to apply
when there are
spatial fluctuations of magnetic field, provided the mean
magnetic field is non-vanishing. When the magnetic field
has zero mean one enters a different regime that is
perturbatively inaccessible from Landau levels;
rather, a different kind of incompressibility may provide
a more suitable starting point for perturbative analysis.
Such a regime can be achieved in two-dimensional systems
with quenched spatial disorder, such as the random
flux (RF) model ${\cal H}_{rf}=(-i \nabla \! + \! {\bf A}/2)^2$
(in gauge $\nabla \! \cdot \! {\bf A}$=0 for our purposes),
and the non-hermitian passive scalar advection-diffusion
model (PS) ${\cal H}_{ps} = \nabla^2 + \nabla \cdot ({\bf v} \cdot)$,
with velocity field ${\bf v} = {\bf A} + {\bf u}$ and vorticity
$\omega={\bf \nabla} \times {\bf A}$, where
$\nabla \! \cdot \! {\bf A}=0$ and $\nabla \! \times \! {\bf u}=0$.

Common to both these models is a component of the form
${\cal K} = {\bf A} \! \cdot \! \nabla$; for example,
${\cal H}_\omega \! = \! \nabla^2/2m \! + \! {\bf A} \! \cdot \!
\nabla \! \rightarrow \! {\cal K}$ as $m \! \to \! \infty$.
As shown in this Letter, incompressibility determines the character of
eigenstates of ${\cal K}$ even when $B$ (or $\omega$) $=-\nabla^2 \phi$
is {\it not} uniform. Models containing ${\cal K}$ are ubiquitous,
and basic features of the spectrum and eigenstates of ${\cal K}$
can be expected to survive perturbation by $-\nabla^2$ and ${\bf A}^2$
or $V$ provided the perturbations are sufficiently weak. 

The quenched random fields of the models fall into two classes.
The vorticity $\omega = \nabla \! \times \! {\bf A}$ in ${\cal H}_{ps}$ is,
like magnetic field, a {\it pseudo}scalar quantity, in contrast to the
scalar $\nabla \! \cdot \! {\bf u}$.
Similarly, the randomness in ${\cal H}_{rf}$ can be decomposed into
parity--odd and -even contributions $i {\bf A} \cdot \nabla$ and ${\bf A}^2/4$. 
This distinction governs the contribution to scattering, and suggests
the study of models, which I call RFI and PSI, that retain only the
parity--odd random components of RF and PS respectively, with 
equations of motion:
$$
(i) \partial_t \psi = \nabla \! \cdot \! {\bf J}_{\psi} 
=  \nabla \! \cdot \! \lbrace (-)\nabla \psi + (i) {\bf A} \psi \rbrace
\eqno(1)
$$
(The factors of $i$ and $-1$ should be chosen appropriately for each model,
but with respect to the symmetry addressed in this paper {\it these factors are
inconsequential}\,.) 

The divergence form of $(1)$ shows explicitly that
RFI and PSI conserve {\it three} currents\cite{Mar98}:
the ordinary probability-current, conserved also when
parity-even fields are present, but
in addition the $\psi$-current ${\bf J}_{\psi}$ generated by $\psi \! \to \! \psi \! + \! C$
and its transpose $\psi^T$-current. When second quantized,
the currents correspond to three non-commuting operators:
the number operator $N_0=\sum_i c^{\dagger}_i c_i$, and
the $\psi$-current operators $c^{\dagger}_0=\sum_i c^{\dagger}_i$ and
$c_0=\sum_i c_i$, the sums being over positions in space.
All three operators commute with $\nabla^2$
and parity-odd operators, but only $N_0$
commutes with potential disorder. The other two fluxes embody
a {\it dynamical} constraint\cite{fluxes} on relaxation
of both the real and imaginary parts of $\psi$: when it is satisfied,
Re$\psi$- and Im$\psi$-fluctuations leaving one region must be
exactly compensated somewhere else;
$V$,${\bf u}$, and ${\bf A}^2/4$ violate the constraint by
introducing uncompensated sources and sinks of Re$\psi$ and Im$\psi$.
When the current ${\bf J}_{\psi}$ and its transpose are conserved,
I say that the coupling to the random field is {\it incompressible}\cite{incom}.

In obtaining RFI from RF by neglecting ${\bf A}^2$,
I have followed ref.[6], where
standard methods\cite{Efetov} are then used to map RFI to
a unitary non-linear sigma model (NLSM), for which {\it all}
eigenfunctions are localized. From the localization of all
eigenfunctions of RFI, it is deduced in ref.[6] that
all eigenfunctions of RF are localized. In contrast, I 
show below that an infinite number of eigenstates of RFI
are extended, which suggests that when ${\bf A}^2$
can be properly treated as a perturbation to RFI, the
corresponding eigenfunctions of RF will be extended.
The conservation laws (1) should be reflected in the correct
NLSM for incompressible disorder, but have no counterparts
in the unitary NLSM, indicating that
the conclusions of ref.[6] may not be universally valid.

Generically, $i{\bf A} \! \cdot \! \nabla$ does not commute with
$-\nabla^2$, and the spectra of these operators
differ qualitatively.
If $i{\cal K} \psi \! = \! E \psi$, then
$i{\cal K} \psi^T \! = \! -E \psi^T$ and
$i{\cal K} \psi^n \! = \! n E \psi^n$,
so the spectrum is symmetric and unbounded. 
It follows that at zero temperature, whereas the ${\bf k}={\bf 0}$ mode of
a free particle is buried at the center of the fermi disc,
for $i {\bf A} \cdot \nabla$ this mode sits instead
at the center of the spectrum, yielding a dirac sea
rather than a fermi sea.

To further study RFI, I at first examine the spectrum and
eigenfunctions of ${\cal K}$ alone, and only later restore $-\nabla^2$.
Eigenfunctions of $i{\cal K}=i\nabla \times \phi \cdot \nabla$
lie on characteristics: the trajectories of $\dot{{\bf x}} \! = \! {\bf A}$,
or equivalently, the streamlines (equipotentials) of $\phi$.
The symmetry of ${\cal K}$ is elucidated
by observing that ${\cal K}\psi$
is just a Poisson bracket, if $\phi$ is viewed as a Hamiltonian
and $y$ as momentum conjugate to $x$:
$ {\cal K}\psi \! =\! \lbrace \psi,\phi \rbrace_{PB} \!
= \! \partial_x \psi \partial_y \phi - \partial_y \psi \partial_x \phi$,
so that the time evolution of $\psi$
is given by $0= d\psi/dt = \partial_t \psi + \lbrace \phi,\psi \rbrace_{PB}$.
Incompressible reparameterizations
are the canonical transformations of $(x,y)$.
${\cal K}$ is covariant
under this non-abelian group of transformations \cite{Sudarshan}.

Under a subgroup of the canonical transformations,
${\cal K}$ is {\it invariant}. This subgroup
represents translations along the streamlines 
of $\phi$; in infinitesimal form ($\eta \ll 1$),
${\bf x} = {\bf {x^{\prime}}} + \eta \nabla \times f(\phi)$,
so that $\phi({\bf x}) = \phi({\bf {x^{\prime}}})$.
These translations are generated by $f(\phi)$, which satisfies
$\nabla \! \times \! \phi \cdot \! \nabla f(\phi) =0$ and represents
an infinite set of conserved charges, one for each streamline.
The Lagrangian corresponding to $\phi(x,p)$ is not
invariant under these translations,
but varies under canonical transformation by a total
time derivative\cite{Sudarshan}, which can lead to
an anomaly\cite{Jackiw}.

Time-independent canonical transformations act
isospectrally on ${\cal K}$; it is instructive to rewrite
the Hamiltonian in action-angle variables.
Take $s$ to be arc-length along $\phi({\bf x})=\phi_0$ and
set $\chi=\int_0^s {ds^{\prime}/{\vert \nabla \phi \vert}}$,
where ${\bf x}(s=0)$ is an arbitrarily chosen point
on the streamline.
Defining the period $\tau(\phi_0) = \oint_{\phi_0} {d \chi }$,
one obtains as variables the area enclosed by the streamline,
$J(\phi_0)=\int_{\phi_0} y \, dx$ (action), and $w=\chi/{\tau}$ (angle).
The eigenstates of
$i{\cal K} = i {\bigl (\tau(\phi_0)\bigr )}^{-1} \partial_w$ 
on equipotential $\phi_0$ are
$$
\psi_{m,\phi_0} = e^{ 2 \pi i m w}
\delta \bigl(J(\phi) - J(\phi_0)\bigr)
\eqno(2)
$$
with eigenvalues $E_{m,\phi_0} = 2 \pi m / \tau(\phi_0)$.
In this local parameterization, the form of the Hamiltonian is
(half) a chiral fermion, yielding the dirac sea obtained above.

It is helpful to examine the simpler case when
$\phi$ is a function of only the coordinate $x$,
where ${\bf x}$ lies on a torus $T^2$
of side $L$; $\phi(x \! + \! L) \! = \! \phi(x)$.
The spectrum is then
$\lbrace (2 \pi m/L) \lbrace \partial_x \phi(x) \rbrace_{x_0}
: 0 \leq x_0 < L \rbrace$.
Zero modes are functions independent of $y$, and
the fourier modes $\exp{(i 2 \pi nx / L)}$, $n$ integer,
constitute a basis of extended eigenstates at $E=0$.
Each non-zero eigenstate
$\delta(x-x_0) f(x) \exp{(i 2\pi m y/L)}$, for integer $m$,
corresponds to a one-dimensional
wave-function, confined transversely to $x_0$ and
extended longitudinally in $y$ around the torus.

To see that the non-zero eigenstates are extended in
the $y$ direction,
one applies a phase twist $\alpha$ at $y=0$. A unitary
transformation by $\exp{i \alpha x/L}$ then yields
$i{\cal K}_{\alpha}=(\partial_x \phi)(i\partial_y + \alpha/L)$. As $\alpha$ varies
from $0$ to $2\pi$, the eigenvalues for any fixed
$x_0$ shift up or down an energy level depending on the sign of
$\partial_x \phi$, whereas the wavefunctions remain
constant. This sensitivity of eigenvalues to perturbation
at the boundary is known as \lq spectral flow,\rq\
and is the hallmark of an anomaly\cite{Jackiw}.

So long as $\phi$ is independent of $y$, the model can
be directly rewritten in fourier basis on an $N \times N$
lattice. Of the $N^2$ modes, $N^2-N$ are
extended in the $y$ direction and localized in the
$x$ direction, and the remaining $N$ zero modes,
the $m=0$ fourier mode on each streamline,
are extended in both the $x$ and $y$ directions.

For generic $\phi(x,y)$, the non-zero eigenstates remain one-dimensional
and confined to the streamlines of $\phi({\bf x})$;
however, they no longer all traverse the torus. One expects
generically two sets of streamlines with opposite orientation
that circulate one of the holes in the torus, but most eigenstates
close without circulating the torus; they are localized.
The zero modes are once again composed of
functions that are constant along the
streamlines, but have arbitrary variation perpendicular
to the streamlines;
$e^{2\pi i n J}$ for integer $n$ constitute
a suitable basis of extended eigenstates at $E=0$.
Eigenstates on streamlines that circulate the torus will
once again shift by an energy level as
the boundary phase twist is increased, but the localized
closed loops will undergo spectral flow only if a phase
twist is imposed somewhere along the loop.

The eigenstates may be further characterized by defining
translation operators $T_{\mu}=e^{\eta \partial_{\mu}}$ for
$\mu=w,J$ corresponding to the pseudomomentum operators
$i\partial_{\mu}$. $T_w$ translates wave functions parallel to 
streamlines; it acts on zero modes as the identity and commutes
with ${\cal K}$, revealing a residual translation invariance
for arbitrary $\phi$ that makes ${\cal K}$ locally one-dimensional.
$T_J$ deforms states in directions {\it perpendicular} to
streamlines: $T_J f(J) = f(J + \eta)$; it does not commute with
${\cal K}$ in general, but does once ${\cal K}$ is projected onto the
$0$ eigenspace, reflecting an enhanced symmetry: the $E=0$ subspace may
be decomposed into eigenstates of $T_J$ and $T_w$ simultaneously.
The deformations represent gapless modes, and are the counterpart
of the area-preserving deformations that leave the 0
mass limit of ${\cal H}_B$ invariant. The symmetries of
the continuum model are fully incorporated into the algebra of 
Poisson brackets, and a faithful
discretization ought to preserve this algebra.

$\nabla^2$ represents a singular perturbation to ${\cal K}$, and
to proceed further ${\cal K}$ must be regularized.
On the lattice, the chain rule, which is needed to show
that ${\cal K}f(\phi)=0$, breaks down for local discretizations,
so that a naive regularization of ${\cal K}$ destroys the conservation laws;
at most a pair of eigenstates with zero energy survive.
A similar problem applies to a naive fourier discretization with momentum cutoff.
Nevertheless, in taking the continuum limit,
the energies of a subset of the modes approach zero;
their eigenfunctions in the discrete theory are smoothly connected
to those of the continuum theory as the UV cutoff is varied
and their identification with zero modes of the continuum model remains.
For this reason, the choice of regularization is not expected to affect 
physical quantities; however, if one needs to identify the
subspace of zero modes explicitly, a symmetry preserving regularization is essential.

The choice of regularization is therefore dictated by a need to
preserve the algebraic properties of the continuum model,
even after truncating the system to a finite number of degrees of freedom.
This requirement is satisfied\cite{Dima} by the
Lie algebras $su(N)$, where $su(N) \! \to \, $sdiff$(T^2)$
as $N \!\! \to \!\! \infty$.
Setting $p=-i \hbar \partial_x$ in $\phi(x,p)$, one obtains
the Heisenberg-Moyal bracket:
$\lbrack \phi(x,-i\hbar \partial_x),\psi(x,-i\hbar \partial_x) \rbrack /
\hbar \!\! \to \!\!
\lbrace \phi(x,p),\psi(x,p)\rbrace_{PB}$
as $\hbar \!\! \to \!\! 0$.
Taking $\hbar=2\pi/N$ for integer $N$
yields, in fourier representation, the algebra $su(N)$
with $N^2-1$ degrees of freedom.
Note that this is the {\it adjoint} representation of
$su(N)$: ${\cal K}\psi=[\phi,\psi]$, the wave-function
is an operator, and
the zero modes of ${\cal K}$ are equivalent under
unitary transformation to the $N-1$ Casimir
elements of $su(N)$ together with the identity.

In fourier space on the torus $[0,1] \! \times \! [0,1]$ 
the Heisenberg-Moyal bracket takes the form:
$$
  [\phi,\psi]_{\bf k}=\sum_{\bf l} {\phi_{{\bf k}-{\bf l}}
  (2\pi/N)\sin((N/2\pi)({\bf k}\times {\bf l})) \psi_{\bf l} }
  \eqno(3)
$$
where ${\bf k},{\bf l}=2\pi (m,n)/N$ , $0 \le m,n < N-1$.
As shown in fig. 1a,
for which $\nabla^2 \phi$ is taken as random and $\delta$-correlated
on distances of order $1/N$, this form correctly reproduces
the overall spectrum obtained via lattice regularization that
is shown in fig. 1b, except in
the neighborhood of $0$ energy, where the $N$ zero modes
appear, separated from the rest of the spectrum by a pseudogap.
Qualitatively, the gap can be thought of as originating
in level repulsion; from eq. (2) it is clear that the 
low-lying states come from the neighborhood of saddle
points of $\phi$.  The scaling of the size of the gap with
$N$ and its dependence on the correlations of $\phi$ can
be estimated by observing that if 
$\phi(\vec{k}) \sim \vert k \vert^{\ell}$ then
$ \nabla \times \phi \cdot \nabla  \sim N^{-(\ell+2)}$;
however, $1/N$ of the spectral weight must be transferred
to the zero modes, yielding a gap of order $N^{-(\ell+3)}$;
$\ell=-2$ in fig. 1.

My continuum arguments indicate that the eigenstates
corresponding to Casimir elements of $su(N)$
should be, in general, extended.
I have confirmed this hypothesis by numerically evaluating 
their Chern number\cite{chern}.

Chern number is computed by  parameterizing
the Hamiltonian by periodic variables $\theta_x,\theta_y$
and integrating the Berry flux over the Brillouin zone.
The degeneracy of ${\cal K}$ at $0$ energy leads to
Berry matrices\cite{Shapere} rather than scalar Berry phases;
to avoid this technical problem, I compute the Chern
number for eigenstates of the operator
${\cal H}=-\nabla^2 + i{\cal K}$.
There is no reason to expect that diffusion, or for
that matter any other non-random, homogeneous coupling,
should localize otherwise delocalized states.
It turns out that an extensive number of states with 
non-zero Chern number is found only within the \lq diffusion band,\rq\
the energy scale set by $-\nabla^2$;
these states span the entire diffusion band.

%\begin{figure*}[btp]
\begin{figure}
%\begin{center}
\begin{picture}(50,-20)

\put(-35,-200){\special{hscale=55 vscale=40 psfile=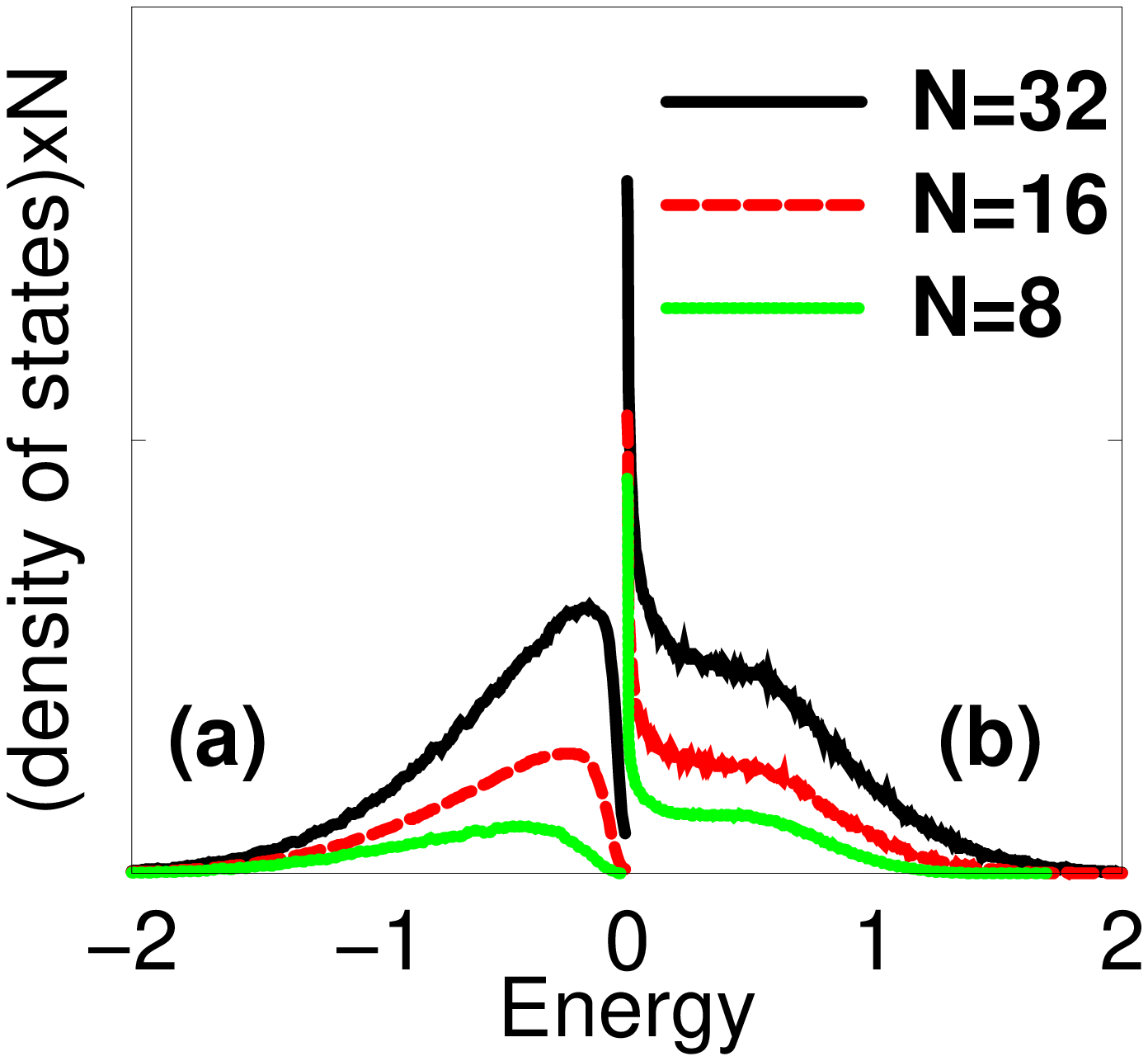}}
\put(-8,-145){\special{hscale=50 vscale=40 psfile=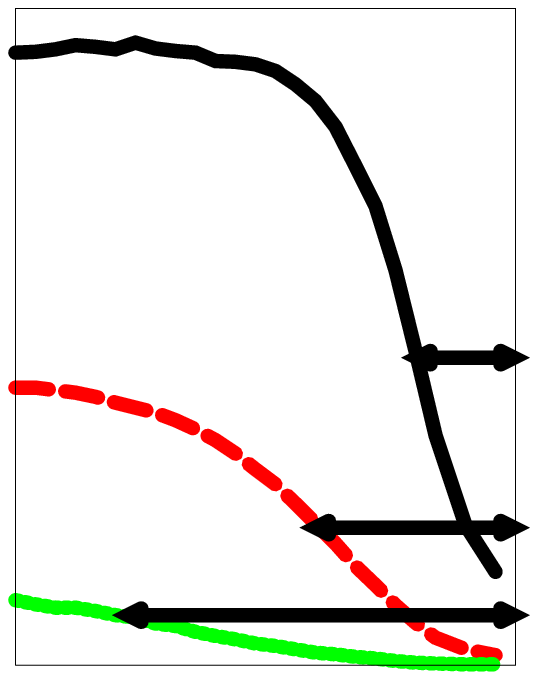}}

\end{picture}
\vskip2.25in
\caption{Density of states of $i{\cal K}$ in (a) $su(N)$ discretization,
and  (b) $N \times N$ lattice discretization, averaged over many realizations
of $\phi$. In (a) the $N$ states
at $E=0$ are suppressed for clarity. The inset expands (a) near $E=0$
to exhibit the scaling of the pseudogap. Except near $E=0$,
almost all states are localized.}

%\end{center}
\label{fig1}
\end{figure}
%\end{figure*}

Fig. 2 shows the fraction of eigenstates within the diffusion
band having non-zero Chern number for a naive lattice discretization
of ${\cal H}$ on an $N \times N$ grid, as a function of $N$,
averaged over many disorder realizations\cite{Mar98}.
The components of ${\bf A}$ were chosen uniformly
on an inteval $[-W,W]$ with $\delta$-correlations in space,
and the compressible part was subtracted.
$\theta_x$ and $\theta_y$ represent phase twists
applied at the boundary of the torus.
The constant asymptote is consistent with a band of extended states in
the thermodynamic limit. 

The inset shows Chern numbers for a typical disorder realization,
discretized with eq. (3) and $N=24$. For this form of ${\cal H}$,  kinetic
energy has been parameterized
as $2\cos(2 \pi m/N + \theta_x) + 2\cos(2 \pi n/N + \theta_y) - 4$,
$ 0 \le m,n \le N-1 $. Nearly every eigenstate in the diffusion
band has a non-zero Chern number; outside the diffusion band,
only isolated energies, corresponding presumably to the
percolating streamlines, show non-zero Chern number. When $i{\cal K}$
is replaced by $V$, all eigenstates 
have zero Chern number.
These numerical results confirm extensive transfer matrix
studies on the lattice\cite{Mar98} for both RFI and PSI.

\begin{figure}
%\begin{center}
\begin{picture}(50,-20)

\put(-30,-200){\special{hscale=53.5 vscale=53.5 psfile=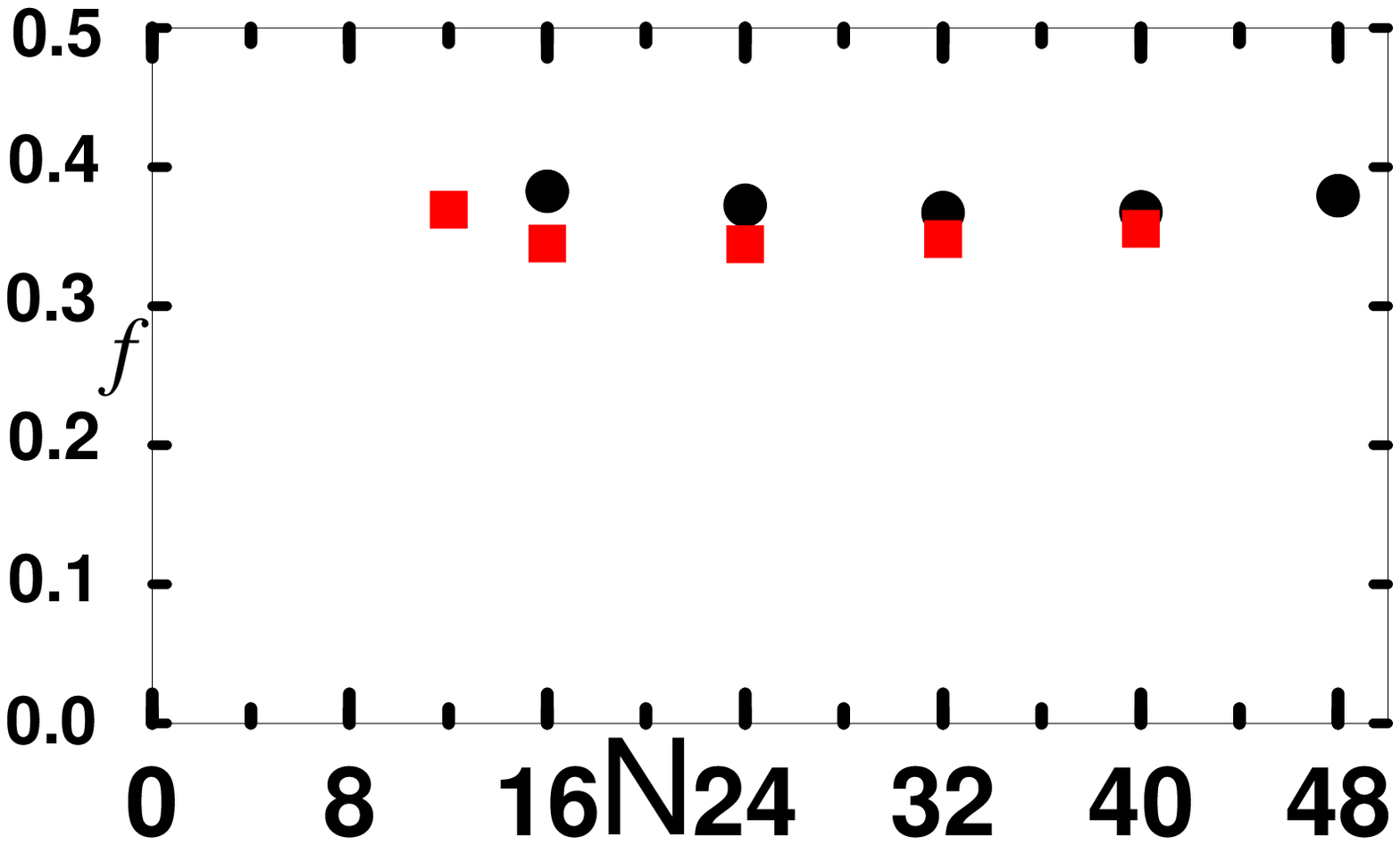}}
\put(45,-167.5){\special{hscale=40 vscale=40 psfile=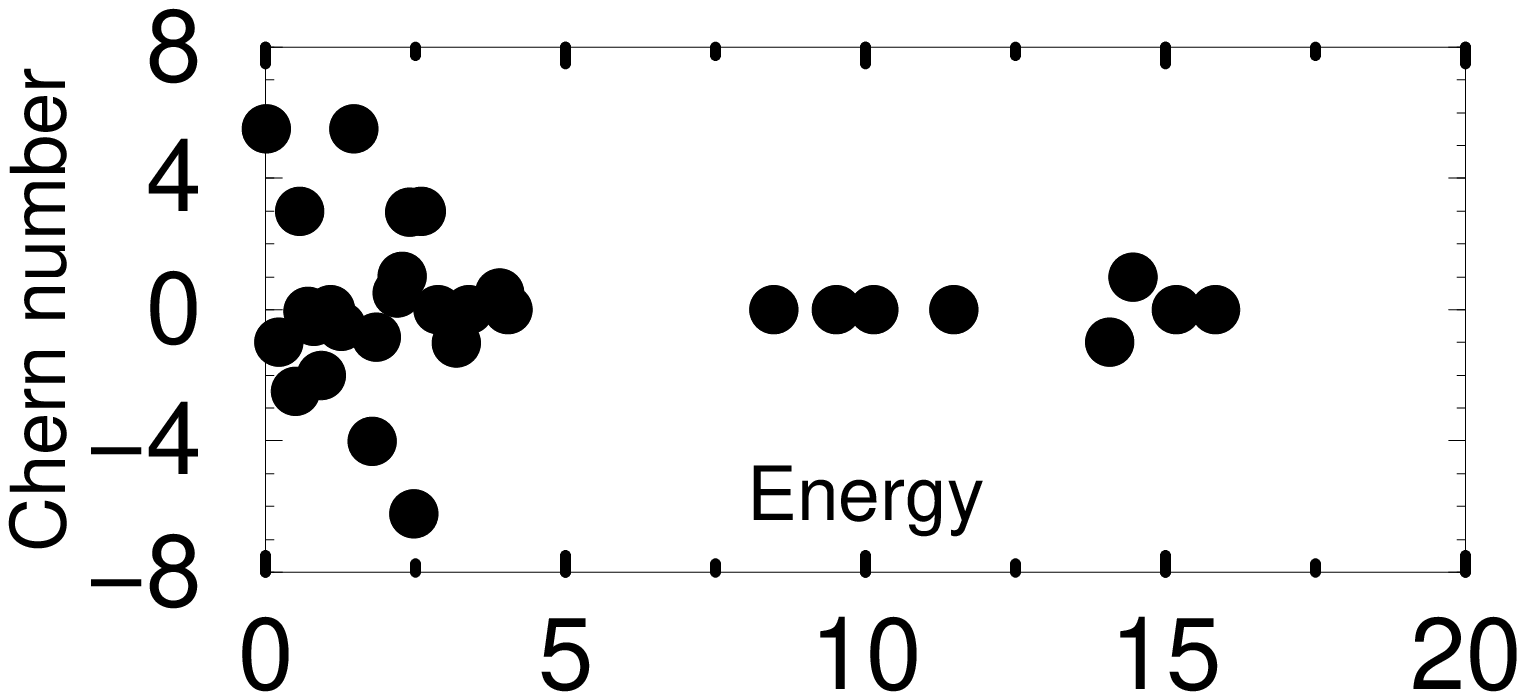}}

\end{picture}
\vskip2.15in
\caption{Fraction $f$ of states within diffusion band having
non-zero Chern number, in lattice discretization of
$-\nabla^2 + i{\cal K}$. The disorder amplitude $W$ (see text)
for circles $45.2$; for squares, $22.6$.
Inset shows (integer) Chern numbers of
eigenstates for a single realization of disorder and
$0 \! < \! E \! < \! 16$, in
$su(N)$ discretization.}
%\end{center}
\label{fig2}
\end{figure}

The phenomena described in this paper originate in
the infinite number of conserved fluxes that, for ${\cal K}$, are
reflected in a huge degeneracy at zero energy. The
charges of the degeneracies produce non-zero Chern numbers\cite{Shapere}
when $i{\cal K}$ is perturbed by $-\nabla^2$, showing
that the existence of extended states does not rely
on the infinite degeneracy itself, but only on the
fact that the perturbation doesn't lead to cancellation
of the charges.

More generally, one might expect that the infinite
number of fluxes, and the symmetry that generates them,
would lead to an anomaly. This point of view applies naturally
to vorticity $\omega$ in a 2d incompressible fluid, where
$\int \! da f(\omega)$
represents the constants of the motion corresponding to the $f(\phi)$
discussed here. Vorticity pumped into a flow at large scales
can be destroyed at small scales only by local action
of $\nabla^2$, creating fluxes of conserved charge in momentum space.
There is no reason to single out a particular charge, such as enstrophy,
for special treatment, as in ref.\cite{Polyakov}.

I have exhibited an extensive fraction of delocalized states
in the diffusion band when $i{\cal K} \gg \nabla^2$, that is,
in the strong disorder limit, whereas arbitrarily weak potential
disorder in $d=2$ localizes all states of $-\nabla^2 + V$.
Barring some intermediate reentrant localized phase,
which seems improbable, it is likely that the states 
will remain extended into the weak disorder regime
$i{\cal K} \ll \nabla^2$. In RG language, one would naively
expect fixed points at $i{\cal K}=0$ and $D\nabla^2=0$,
both of which display an infinite number of extended eigenstates.

The $i{\cal K} \ll \nabla^2$ regime could be achieved if the
random magnetic field $B$ had strong anti-ferromagnetic correlations
so that $\phi \sim 1$ and consequently ${\bf A} \sim k$.
The center of the band sets a momentum scale $k_f$,
and the order of magnitude of the terms in RF may be estimated as
$\nabla^2 \sim k_f^2$,$\, i{\cal K} \sim k_f k$, and ${\bf A}^2 \sim k^2$,
suggesting that it may be reasonable to incorporate the effect of
the compressible disorder ${\bf A}^2$ as a perturbation to RFI.

I conclude by addressing the effect of compressible disorder on RFI.
While eigenstates of $\nabla^2$ with opposite momentum are degenerate,
as of course are $0$ modes of $i{\cal K}$, these degeneracies are broken
for the extended eigenstates of $-\nabla^2 + i{\cal K}$; consequently,
those eigenstates are chiral and small compressible disorder should not
localize them. Rather, it can be expected to compete with incompressible
disorder, with its localizing effects most strongly felt near the band edges.
For example, when $A^2$ is negligible compared to $-\nabla^2 + i{\cal K}$
then the entire diffusion band (or blob, for PSI) should
be extended. When $\phi \sim 1/{\vert {\bf k} \vert}$, the critical
energy ought to depend on the amplitude of the vector potential randomess;
as the relative strength of $A^2$ increases, the extended regime should contract
to the point of particle-hole symmetry\cite{mw}. For these reasons, one
might imagine that the divergent density of states recently found at the band
center of a random-flux model\cite{mw} reflects the dirac sea I have
identified above.

I am grateful to P.~Weichman, C.~Mudry and B.~Bialek for extensive comments
on the manuscript, to D.~Huse for observations about spatially anti--correlated
flux distributions, to M.~Douglas for remarks on canonical
transformations, and to R.~Lifshits and M.~Cross for
support and encouragement. The numerics were carried out on the CACR parallel
computer system at Caltech.


\begin{references}
\bibitem{fluxes} Conservation of the $\psi$-currents does
not mean simply that ${\cal A}$ vanishes for a constant
wave function; consider
$ {\cal A} = i\sum_{i,j} a_{ij}
 \lbrace c^{\dagger}_i c_j - c^{\dagger}_j c_i \rbrace $,
$a_{ij}$ spatially random.
\bibitem{incom}
The term \lq incompressible\rq\
is justified since reparameterization
by an arbitrary incompressible coordinate transformation
preserves all three conserved currents:
%
$$
	(1/{\sqrt g}) (\partial_i {\sqrt g} g^{ij} \partial_j
	+ i \epsilon^{ij} \partial_i \phi \partial_j)
$$
%
can be written in divergence form, provided $g=det(g_{ij})$ is
constant, $g_{ij}$ the metric tensor.
\end{references}
\end{document}